\documentclass{iaus}
\usepackage{graphicx}

\title[DIBs vs. Known Gas-Phase Species] 
{Diffuse Interstellar Bands:  How Are They Related to Known Gas-Phase Constituents of the ISM?}

\author[D. E. Welty]   
{Daniel E. Welty}

\affiliation{University of Chicago, Astronomy \& Astrophysics Center, \\ 5640 S. Ellis Ave., Chicago, IL 60637, USA \\ email: {\tt dwelty@oddjob.uchicago.edu}}

\pubyear{2013}
\volume{297}  
\pagerange{xxx--xxx}
\jname{The Diffuse Interstellar Bands}
\editors{J. Cami \& N. L. J. Cox, eds.}

\begin{document}

\maketitle

\begin{abstract}
In this brief review of recent work relating the DIBs to other gas-phase constituents of the ISM, we explore correlations between DIB equivalent widths and the column densities of various atomic and molecular species, drawn from a large database constructed for that purpose.  
The tightness and slopes of the correlations can provide information on how the DIBs might be related to those species (physically, chemically, spatially) and on various properties of the DIB carriers.
Deviations from the mean relationships can reveal dependences of DIB strengths on other parameters, regional variations in DIB behavior, and individual sight lines where unusual environmental conditions affect the DIBs.  
Variations in DIB profiles (e.g., wings,  substructure) and relative strengths may be related to differences in physical conditions inferred from atomic and/or molecular absorption lines.

\end{abstract}

\section{Introduction}

High-resolution optical/UV spectra of interstellar atomic and molecular absorption lines have revealed complex spatial and velocity structure in the predominantly neutral component of the Galactic ISM.  
That neutral component can exhibit a range in physical conditions -- from warm, low-density atomic gas to cold, dense molecular gas.  
Where do the DIBs arise, and what can their behavior tell us about the conditions where their carriers reside?  
In the following sections, we briefly note some issues that should be kept in mind when comparing the DIBs with other constituents of the ISM, we revisit correlations between DIB strengths and the abundances of various atomic and molecular species that have been discussed in previous studies, and we describe several cases where the DIBs appear to be affected by unusual environmental conditions.

\begin{figure}[t!]
\begin{center}
\includegraphics[width=5.0in]{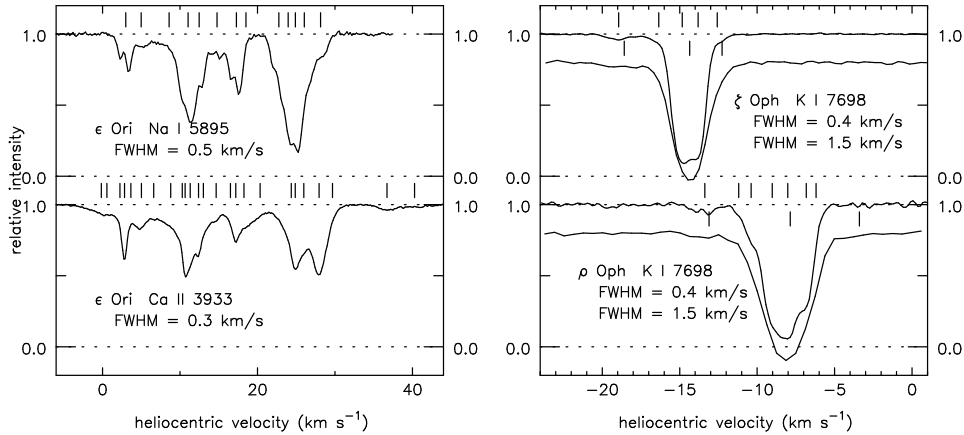} 
\caption{Complex interstellar velocity structure seen in very high resolution (FWHM $\le$ 0.5 km~s$^{-1}$) spectra. 
Toward $\epsilon$ Ori ({\it left}), many components are seen for both Na~{\sc i} and Ca~{\sc ii}; the Na~{\sc i}/Ca~{\sc ii} ratio can vary significantly.
Toward $\zeta$ Oph and $\rho$ Oph ({\it right}), the apparently ``single'' components seen in K~{\sc i} at a resolution of 1.5 km~s$^{-1}$ ({\it lower spectra}) are shown at higher resolution ({\it upper}) to be blends of several narrow ($b$ $\le$ 0.6 km~s$^{-1}$), closely-spaced components.}
\label{fig:vstruct}
\end{center}
\end{figure}

\section{Complex Structure and Processes in the ISM}

{\bf Small-scale spatial structure:} 
Sub-pc spatial structure in predominantly neutral interstellar clouds has been inferred from variations in H~{\sc i} absorption toward pulsars and resolved extragalactic sources and from variations in Na~{\sc i} absorption toward binary/multiple stellar systems and stars with large proper motions (e.g., \cite[Lauroesch 2007]{Laur07}).
While the variations in H~{\sc i} have been ascribed to dense filamentary or sheet-like structures (\cite[Heiles 1997]{Heil07}), the variations in Na~{\sc i} (and other trace neutral species) may instead be due to small-scale fluctuations in the local density $n_{\rm H}$ and/or the electron density $n_{\rm e}$ (\cite[Welty 2007]{Welty07}).
Similar small-scale variations in DIB strengths have been noted in a few cases; toward the $\rho$ Oph system, stronger variations are seen for some of the DIBs than for the atomic and molecular species -- suggesting that the abundances of those DIBs can be quite sensitive to the local physical conditions (\cite[Cordiner et al. 2006]{Cord06}, 2013).

{\bf Complex velocity structure:} 
Complex interstellar velocity structure -- with many narrow, closely-spaced components -- is present in many Galactic sight lines (e.g., \cite[Price et al. 2001]{Price01}; \cite[Welty \& Hobbs 2001]{WH01}).
Sight lines that appear to have a single dominant component even at resolutions of 1.5 km~s$^{-1}$ (higher than used for most DIB observations) often exhibit finer structure when observed at higher resolution (Fig.~\ref{fig:vstruct}, {\it right}).
The individual components can have very different relative abundances of the various atomic and molecular species (Fig.~\ref{fig:vstruct}, {\it left}) -- presumably reflecting differences in local physical conditions.
The small $b$-values for trace neutral atomic and molecular species (e.g., median $b$ $<$ 0.7 km~s$^{-1}$ for K~{\sc i}) imply that saturation can be an issue even for relatively weak lines.
The somewhat larger $b$-values found for Ca~{\sc ii} suggest that Ca~{\sc ii} (and various dominant singly ionized species) may be more broadly distributed -- even within a single ``cloud'' (e.g., \cite[Welty et al. 1996]{WMH96}).
While velocity structure corresponding to widely separated component groups can be seen for DIBs in some sight lines (e.g., \cite[Weselak et al. 2010]{Wes10}), the more complex structure on much smaller velocity scales is not discernible in even the narrowest DIBs (which have FWHM $\sim$ 15--20 km~s$^{-1}$).

{\bf Ionization equilibrium:} 
Until fairly recently, the abundances of trace ions (e.g., C~{\sc i}, Na~{\sc i}, K~{\sc i}) in predominantly neutral clouds were generally considered to reflect a simple balance between photoionization (due to the ambient UV radiation field) and radiative recombination (with dielectronic recombination for some species in warmer gas).
Observations of H$_3^+$ absorption (e.g., \cite[Indriolo \& McCall 2012]{Ind12}), however, suggest that the cosmic-ray ionization rate is generally higher than previously assumed (though somewhat variable) -- with significant effects on both the ionization and chemistry in diffuse clouds.
Moreover, charge exchange between dominant singly ionized gas-phase atomic species and neutral or negatively charged small grains (or large molecules) can significantly enhance the abundances of some of the corresponding trace neutral species (\cite[Weingartner \& Draine 2001]{Wein01}; \cite[Liszt 2003]{Liszt03}).
Inclusion of this ``grain-assisted recombination'' still does not yield complete agreement among the $n_{\rm e}$ inferred from various X~{\sc i}/X~{\sc ii} ratios, however (\cite[Welty et al. 2003]{WHM03}).
Values of $n_{\rm H}$ inferred from those $n_{\rm e}$ (assuming most of the electrons come from ionization of carbon) must be considered to be very uncertain.

{\bf Depletions:} 
Elements deficient in the gas-phase (relative to Solar abundances) are thought to be ``depleted'' onto dust grains that are generally coextensive with the gas.
The severity of the depletions can vary considerably -- for different elements, different sight lines, and different parcels of gas along a given sight line.
While the depletions of the more refractory elements (Fe, Ni, Ti) can range from factors of 2--3 in very lightly reddened sight lines to factors $>$300 in heavily reddened sight lines containing significant amounts of H$_2$, the depletions of the more volatile elements (C, N, O, Kr) generally do not exceed a factor of 3. 
Using a compilation of abundances for 17 elements in 243 Galactic sight lines, \cite[Jenkins (2009)]{Jenk09} defined a depletion index (F*) for each sight line (assuming consistent relationships between the depletions of the various elements) and explored the implied changes in dust composition with F*.
There may, however, be some regional variations in the depletion patterns -- perhaps in Sco-Oph, and more markedly in the lower metallicity Magellanic Clouds (\cite[Welty \& Crowther 2010]{WC10}). 
And while the depletions of the refractory elements are expected to be increasingly severe in denser regions, those severe depletions can be masked in integrated sight line measures by contributions from more diffuse gas (with much milder depletions), and the expected dependence of the depletion on local density is not apparent in the existing (small) sample (\cite[Welty \& Crowther 2010]{WC10}).
Could depletion of the DIB carriers (or of any refractory constituents of the carriers) contribute to the weakening of the DIBs in higher density gas?

{\bf Correlations:} 
While we would like to compare various constituents of the ISM on an individual component basis, such detailed comparisons generally are not possible -- e.g., for H, H$_2$, $E(B-V)$, and the DIBs (as well as for most other quantities not derived from high resolution spectra).
Nonetheless, comparisons of integrated sight line quantities can help to reveal the physical, spatial, and/or chemical relationships between those quantities (with appropriate caveats).
We will use DIB equivalent widths and the column densities of atomic and molecular species -- which give direct measures of the amounts of each constituent; using log-log plots allows non-linear relationships to be more easily discerned and evaluated.
As most ISM tracers will correlate with each other to some degree, some measure of the ``goodness'' of the correlation is needed -- but even ``good'' correlations do not necessarily imply direct association.
The slopes of the ``best-fit'' lines can give clues to relevant physical relationships/processes and relative spatial distributions; examination of the residuals and ``discrepant'' points can reveal trends with other quantities, regional variations, and cases where extreme environmental conditions affect the DIBs.
Comparison of appropriate ratios can (largely) remove the common dependence on the total amount of interstellar material present and can aid in disentangling the multiple factors affecting the DIB strengths.

\begin{figure}[t!]
\begin{center}
\includegraphics[width=5.0in]{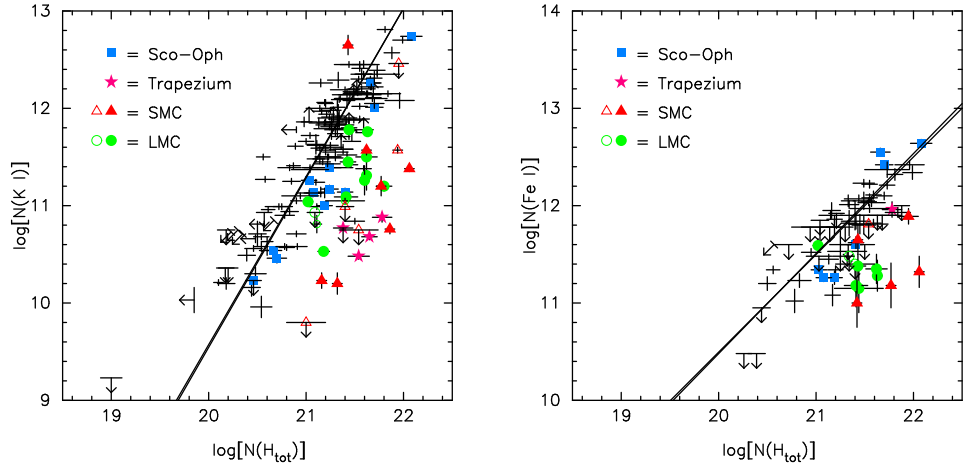}
\caption{Column densities of trace neutral species K~{\sc i} and Fe~{\sc i} vs. $N$(H$_{\rm tot}$).
For most Galactic sight lines, K~{\sc i} follows a nearly quadratic relationship with H$_{\rm tot}$; the relationship for Fe~{\sc i}, however, is roughly linear.
Systematic deviations from the general trends may be noted for several regions.
Note:  the correlations shown in this review are drawn from a database of interstellar quantities for $\sim$480 Galactic and $\sim$290 Magellanic Clouds sight lines ($\sim$180 with equivalent widths for several DIBs).
Fits to the relationships (log Y vs. log X) do not include the Sco-Oph, Trapezium, or Magellanic Clouds sight lines.}
\label{fig:kfe_htot}
\end{center}
\end{figure}

Figure~\ref{fig:kfe_htot} shows the relationships between the column densities of two trace neutral species (K~{\sc i}, Fe~{\sc i}) and the total hydrogen column density $N$(H$_{\rm tot}$) = $N$(H) + 2$N$(H$_2$).
For most Galactic sight lines, K~{\sc i} exhibits a fairly tight relationship with H$_{\rm tot}$, with slope $\sim$ 1.7 -- close to the quadratic relationship expected from considerations of ionization equilibrium (\cite[Welty \& Hobbs 2001]{WH01}).
Enhanced radiation fields may be responsible for the somewhat lower $N$(K~{\sc i}) seen for some sight lines in Sco-Oph, and for the much lower values seen toward the stars in the Orion Trapezium region.
A combination of lower metallicity, enhanced radiation fields, and (perhaps) reduced grain-assisted recombination may account for the lower $N$(K~{\sc i}) in the LMC and SMC (Welty \& Crowther, in prep.). 
The slope for Fe~{\sc i} vs. H$_{\rm tot}$ is smaller -- close to 1.0 -- reflecting the increasingly severe depletion of Fe in the higher column density sight lines (\cite[Welty et al. 2003]{WHM03}).

Similar comparisons among the column densities of the diatomic and triatomic molecular species observable in the optical/UV indicate both differences in the degree of correlation and systematic trends for the correlation slopes vs. H$_2$ -- which may be understood in terms of both the chemical networks linking the molecules and their relative spatial distributions within diffuse molecular clouds (\cite[Pan et al. 2005]{Pan05}; Welty et al., in prep.).
Both CH and OH appear to be formed relatively early (once H$_2$ is present), and have slopes of order 1.0--1.2 vs. H$_2$.
So-called ``second-'' and ``third-generation'' species (C$_2$, C$_3$; CN, CO), which require the existence of other precursor molecules for their formation and which typically trace the denser parts of the clouds, exhibit progressively steeper relationships with H$_2$.
For $^{12}$CO, \cite[Sheffer et al. (2008)]{Shef08} have noted a further steepening of the relationship vs. H$_2$ which marks the onset of $^{12}$CO self-shielding.
(For more detailed discussions of interstellar chemistry and the DIBs, see the contributions in these proceedings by Liszt, Wakelam, Roueff, and Federman.)

\section{DIBs vs. Known Atomic and Molecular Species}

{\bf General surveys:} 
In one early, extensive study, \cite[Herbig (1993)]{Herb93} compared the strengths of the $\lambda$5780.5 and $\lambda$5797.1 DIBs with a number of ISM tracers, for a set of 93 sight lines sampling a variety of environments.
Those comparisons showed $W$(5780.5) to be well correlated with $N$(H), with slope $\sim$ 1.35 (and several ``discrepant'' sight lines); significantly, there was no residual or secondary correlation with $N$(H$_2$).
Both DIBs exhibited roughly linear relationships with $E(B-V)$ and somewhat shallower relationships (slopes $\sim$ 0.7) with the column densities of Na~{\sc i}, K~{\sc i}, and C~{\sc i}; neither showed any dependence on $N$(CH$^+$) or the depletion of Ti (tracers of non-thermal chemistry and grain processing, respectively).
Herbig concluded that the carrier of the DIBs ``appears to be a free neutral, gaseous species \ldots (with an) ionization/dissociation threshold $>$ 5.1 eV.''

\begin{figure}[t!]
\begin{center}
\includegraphics[width=5.0in]{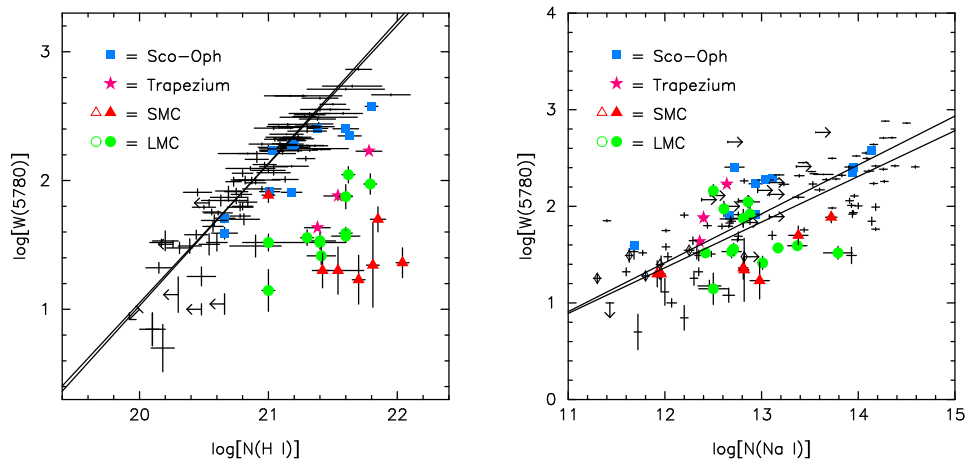}
\caption{Equivalent width of $\lambda$5780.5 DIB vs. $N$(H) ({\it left}) and $N$(Na~{\sc i}) ({\it right}).
For most Galactic sight lines, the $\lambda$5780.5 DIB follows a fairly tight, nearly linear relationship with H, but can be weaker in some regions.
For $W$(5780.5) vs. $N$(Na~{\sc i}), the mean relationship is shallower (slope $\sim$ 0.5), with more scatter (but less regional variation).}
\label{fig:d5780}
\end{center}
\end{figure}

Figure~\ref{fig:d5780} revisits the relationships between the $\lambda$5780.5 DIB and H and Na~{\sc i}, for our larger sample.
As before, there is a good correlation ($r$ $\sim$ 0.90) with $N$(H), but with smaller slope ($\sim$ 1.1).
As for K~{\sc i} (Fig.~\ref{fig:kfe_htot}), the $\lambda$5780.5 DIB is slightly weaker in the Sco-Oph sight lines and much weaker in the Trapezium sight lines -- likely due to enhanced radiation fields.
The general weakness of the $\lambda$5780.5 DIB in the LMC and SMC likely reflects both the lower metallicities and stronger radiation fields; differences in cloud structure (in low-metallicity systems) may also contribute (\cite[Cox et al. 2006]{Cox06}, 2007; \cite[Welty et al. 2006]{Welty06}).
The relationship between $W$(5780.5) and $N$(Na~{\sc i}) is also shallower than found by Herbig (slope $\sim$ 0.5), with larger scatter but less regional variation than for that DIB vs. H.
By analogy with K~{\sc i} and Fe~{\sc i} vs. H$_{\rm tot}$, the carrier of the $\lambda$5780.5 DIB could be either a dominant species (in H~{\sc i} gas) or a trace species which can be depleted/altered/destroyed both at higher $N$(H) and in stronger radiation fields.

Kre{\l}owski, Galazutdinov, Weselak, and collaborators have noted differences in the behavior of several DIBs with respect to some atomic and molecular species.
The DIBs at 5780.5, 5797.1, 5849.8, and 6613.6 \AA\ appear to be better correlated with K~{\sc i} than with Ca~{\sc ii} (\cite[Galazutdinov et al. 2004]{Gala04}); the correlations with K~{\sc i} are better for the narrower DIBs ($\lambda$5797.1, $\lambda$6379.3) than for the broader DIBs ($\lambda$5780.5, $\lambda$6283.8) (\cite[Kre{\l}owski et al. 1998]{Krel98}).
The ratio $W$(5797.1)/$W$(5780.5) appears to be correlated with $N$(CH)/$E(B-V)$ and with the molecular fraction $f$(H$_2$) = 2$N$(H$_2$)/$N$(H$_{\rm tot}$), less correlated with $N$(CN)/$E(B-V)$, and not correlated with $W$(CH$^+$)/$E(B-V)$ (\cite[Kre{\l}owski et al. 1999]{Krel99}; \cite[Weselak et al. 2004]{Wes04}, 2008) -- suggesting that the $\lambda$5797.1 DIB traces somewhat denser gas than the $\lambda$5780.5 DIB, but not the denser regions where most of the CN is located.

\begin{figure}[t!]
\begin{center}
\includegraphics[width=5.0in]{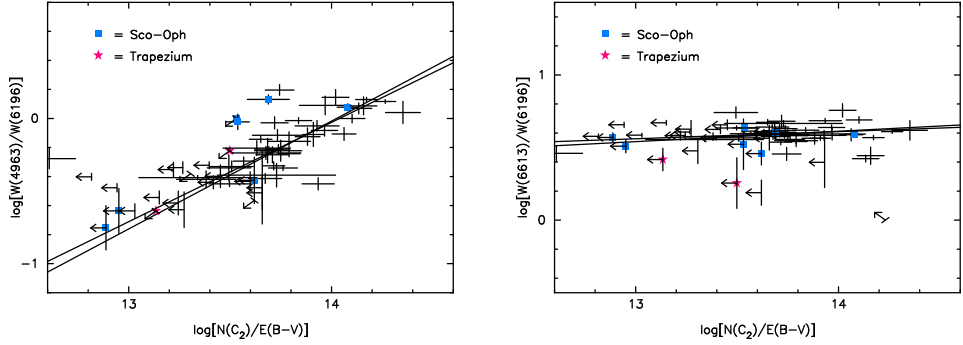}
\caption{Normalized equivalent widths of $\lambda$4963.9 and $\lambda$6613.6 DIBs vs. $N$(C$_2$)/$E(B-V)$.
The $\lambda$4963.9 DIB (a ``C$_2$-DIB'') is generally enhanced when C$_2$ is enhanced; the $\lambda$6613.6 DIB shows no such trend.}
\label{fig:c2dibs}
\end{center}
\end{figure}

\begin{figure}[t!]
\begin{center}
\includegraphics[width=5.0in]{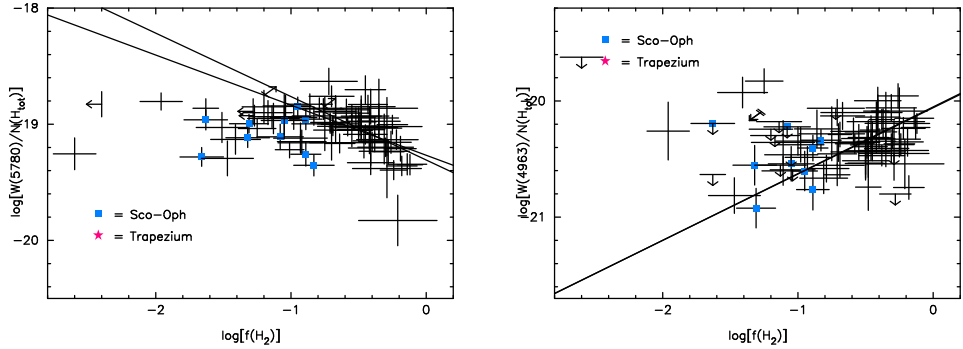}
\caption{Normalized equivalent widths of $\lambda$5780.5 and $\lambda$4963.9 DIBs vs. molecular fraction $f$(H$_2$).
For $f$(H$_2$) $>$ 0.1, the $\lambda$5780.5 DIB weakens with increasing $f$(H$_2$) -- suggesting that it traces mostly atomic gas.
The $\lambda$4963.9 DIB (a ``C$_2$-DIB''), however, seems to strengthen with increasing $f$(H$_2$).
(Note: the fits are for $f$(H$_2$) $\ge$ 0.1 only.)}
\label{fig:dibs_fh2}
\end{center}
\end{figure}

{\bf C$_2$-DIBs:} 
While the $\lambda$5780.5 and $\lambda$5797.1 DIBs seem to trace relatively diffuse, primarily atomic gas, \cite[Thorburn et al. (2003)]{Thor03} identified a small set of relatively weak DIBs which appear to be associated with somewhat denser gas traced by C$_2$.
From a sample of 21 isolated DIBs in high-S/N spectra of 53 sight lines, seven DIBs exhibited trends in the residuals [relative to the mean relationship between $W$(DIB) and $E(B-V)$] vs. $N$(C$_2$)/$E(B-V)$ -- i.e., they seem to be enhanced when C$_2$ is enhanced; 11 other such ``C$_2$-DIBs'' were subsequently identified.
Figure~\ref{fig:c2dibs} shows the difference in behavior for the normalized equivalent widths of the $\lambda$4963.9 and $\lambda$6613.6 DIBs; in both cases, the $\lambda$6196.0 DIB (which shows no residual correlation with the C$_2$ abundance) is used for normalization.
The $\lambda$4963.9 DIB (among the strongest of the C$_2$-DIBs), exhibits a clear correlation with $N$(C$_2$)/$E(B-V)$; the $\lambda$6613.6 DIB (not a C$_2$-DIB) shows no such relationship.
Similar trends are seen vs. $N$(CN)/$E(B-V)$ and (less clearly) vs. $N$(CH)/$E(B-V)$.
All of the 18 identified C$_2$-DIBs are narrow (FWHM $<$ 1 \AA); intriguingly, there are several pairs of C$_2$-DIBs with differences in wavenumber $\Delta\sigma$ $\sim$ 20 cm$^{-1}$, suggestive of a spin-orbit interaction in a linear molecule.
Figure~\ref{fig:dibs_fh2} gives another indication of the relationship between the C$_2$-DIBs and the diffuse molecular gas:  the normalized strength of the $\lambda$4963.9 DIB [$W$(4963.9)/$N$(H$_{\rm tot}$)] appears to increase with increasing molecular fraction [for $f$(H$_2$) $>$ 0.1, where H$_2$ becomes significant], while $W$(5780.5)/$N$(H$_{\rm tot}$) (representative of the majority non-C$_2$-DIBs) declines over that range in $f$(H$_2$).

{\bf New surveys:} 
\cite[Friedman et al. (2011)]{Fried11} investigated the correlations between eight of the stronger DIBs and H, H$_2$, and $E(B-V)$, for a set of 133 sight lines.
All eight of the DIBs are better correlated with H than with H$_2$ (when several sight lines with weak $\lambda$5780.5 DIBs are excluded), and six of the eight are better correlated with H than with $E(B-V)$ -- suggesting that (like $\lambda$5780.5) those DIBs arise in relatively diffuse, predominantly atomic gas.
In Table~\ref{tab1}, we give the linear correlation coefficients ($r$) and best-fit correlation slopes for those eight DIBs (plus the $\lambda$4963.9 C$_2$-DIB) vs. H, H$_2$, CH, H$_{\rm tot}$, $E(B-V)$, Na~{\sc i}, K~{\sc i}, and Ca~{\sc i} (for the full Friedman et al. sample); the DIBs are ordered by (generally) increasing slope.
Several additional trends may be noted:
(1) For six of the nine DIBs, the best correlations are with H$_{\rm tot}$; for eight of the nine, the poorest correlations are with Ca~{\sc ii} ($r$ = 0.21--0.57).
(2) All nine DIBs exhibit similar correlations with $E(B-V)$, with $r$ = 0.79--0.85; curiously, the correlations with the total visual extinction $A_{\rm V}$ are somewhat poorer ($r$ = 0.58--0.74).
Slightly stronger correlations have been found between the $\lambda$5780.5 and $\lambda$6283.8 DIBs and $E(B-V)$ toward a sample of somewhat cooler stars (\cite[Raimond et al. 2012]{Rai12}).
(3) The broader DIBs tend to have smaller slopes than the narrower DIBs -- in most cases, the $\lambda$6283.8 DIB (FWHM $\sim$ 4.77 \AA) has the smallest slope and the $\lambda$6613.6 DIB (FWHM $\sim$ 0.93 \AA) has the largest slope. 
(4) For the Trapezium sight lines, the $\lambda$6283.8 DIB has nearly ``normal'' strength vs. $N$(H$_{\rm tot}$), while the $\lambda$6613.6 DIB exhibits the largest deficiencies.
(5) The narrower DIBs tend to be better correlated with the trace neutral species Na~{\sc i} and K~{\sc i} than the broader DIBs.
By analogy with the similar trends noted for simple molecular species, the correlation slopes may provide clues to the relative spatial distributions (and the physical/chemical properties) of the DIB carriers, and ratios of DIBs exhibiting differences in behavior [e.g., $W$(6613.6)/$W$(6283.8)] may provide additional diagnostics for cloud physical conditions.
Such comparisons will be performed for the full set of DIBs listed in our two atlases (\cite[Hobbs et al. 2008]{Hob08}, 2009).

\begin{table}[t!]
\begin{center}
\caption{Correlations:  nine DIBs vs. H, H$_2$, CH, H$_{\rm tot}$, $E(B-V)$, Na~{\sc i}, K~{\sc i}, Ca~{\sc i}.}
\label{tab1}
{\scriptsize
\begin{tabular}{cccccccccc}
\hline 
DIB   & FWHM&      $N$(H)   &    $N$(H$_2$) &  $N$(CH) &$N$(H$_{\rm tot}$)&     $E(B-V)$&    $N$(Na~{\sc i})  &     $N$(K~{\sc i})  & $N$(Ca~{\sc i})\\
      &(\AA)&      $r$/slope&      $r$/slope&  $r$/slope&      $r$/slope&      $r$/slope&      $r$/slope&      $r$/slope& $r$/slope\\
\hline
6283.8& 4.77&{\bf 0.87}/0.86&      0.46/0.25& 0.47/0.41 &{\bf 0.82}/0.70&{\bf 0.82}/0.82&      0.71/0.39&      0.67/0.39& 0.71/0.70\\
5487.7& 5.20&      0.60/0.93&      0.47/0.29& 0.57/0.49 &      0.72/0.76&      0.79/0.90&      0.56/0.49&      0.58/0.41& 0.62/0.75\\
6204.5& 4.87&{\bf 0.84}/0.93&      0.60/0.29& 0.59/0.49 &{\bf 0.85}/0.79&{\bf 0.83}/0.89&      0.64/0.40&      0.55/0.42& 0.75/0.75\\
5705.1& 2.58&      0.73/1.01&      0.56/0.41& 0.67/0.49 &{\bf 0.82}/0.90&{\bf 0.80}/0.82&      0.69/0.44&      0.68/0.42& 0.60/0.70\\
5780.5& 2.11&{\bf 0.90}/1.05&      0.65/0.34& 0.64/0.48 &{\bf 0.92}/0.90&{\bf 0.82}/0.94&      0.78/0.44&      0.71/0.48& 0.75/0.87\\
6196.0& 0.42&      0.79/1.04&      0.74/0.40& 0.72/0.55 &{\bf 0.92}/0.89&{\bf 0.85}/0.95&{\bf 0.89}/0.51&{\bf 0.84}/0.52& 0.79/0.71\\
5797.1& 0.77&      0.72/1.05&      0.79/0.41& 0.77/0.61 &{\bf 0.91}/0.94&{\bf 0.84}/0.99&{\bf 0.87}/0.56&{\bf 0.84}/0.57& 0.76/0.77\\
4963.9& 0.62&      0.62/1.04&      0.53/0.43& 0.74/0.67 &      0.77/1.09&      0.79/1.07&{\bf 0.80}/0.83&{\bf 0.84}/0.70& 0.62/0.73\\
6613.6& 0.93&      0.77/1.40&{\bf 0.80}/0.54& 0.78/0.63 &{\bf 0.93}/1.19&{\bf 0.83}/1.20&{\bf 0.88}/0.67&{\bf 0.85}/0.72& 0.78/0.94\\
\hline
\end{tabular}
}
\end{center}
\vspace{1mm}
\scriptsize{
 {\it Notes:}
Linear correlation coefficients ($r$) and slopes are for log[$W$(DIB)] vs. log[x].
DIBs are ordered by (generally) increasing slope.
Bold type for $r$ values identifies relationships with $r$ $\ge$ 0.80.
Based on DIB equivalent widths from \cite[Thorburn et al. (2003)]{Thor03} and \cite[Friedman et al. (2011)]{Fried11}.}\\
\end{table}

\section{Regional Differences / Individual Sight Lines}

{\bf Sco-Oph:} 
As noted above, Na~{\sc i}, K~{\sc i}, and some of the DIBs are relatively weak for some sight lines in the Sco-Oph region (Figs.~\ref{fig:kfe_htot} and \ref{fig:d5780}).
Using optical spectra of 89 sight lines in Upper Scorpius, \cite[Vos et al. (2012)]{Vos12} measured equivalent widths for five DIBs and the strongest lines of CH, CH$^+$, CN, K~{\sc i}, and Ca~{\sc i}. 
A simple radiative transfer model designed to reproduce the observed CH and CN yielded estimates for the local UV radiation field strength ($I_{\rm UV}$).
Vos et al. found a bimodal distribution for the $W$(5797.1)/$W$(5780.5) ratio, with a primary peak at 0.20$\pm$0.05; many of the higher values ($\ge$ 0.3) are found for sight lines sampling the ``denser'' gas and dust near $\rho$ Oph.
Higher $W$(5797.1)/$W$(5780.5) ratios were found only for lower inferred $I_{\rm UV}$ -- generally supporting the long-held view that the $\lambda$5797.1 DIB is more sensitive to the radiation field and thus traces somewhat denser, more shielded regions than the $\lambda$5780.5 DIB.

{\bf Herschel 36:} 
Optical/UV spectra of the heavily reddened star Herschel 36 have revealed several very unusual features:  absorption from rotationally excited CH and CH$^+$, absorption from vibrationally excited H$_2$, and strong, extended redward wings on some of the stronger DIBs (\cite[Dahlstrom et al. 2013]{Dahl13}; \cite[Oka et al. 2013]{Oka13}; see contributions in these proceedings by York and Oka).
Both K~{\sc i} and the DIBs are weak (relative to H and H$_{\rm tot}$) in the material near Herschel 36 where those unusual features arise -- suggestive of an enhanced UV radiation field; the upper limits on H$_2$ and CN suggest that the strong CH and CH$^+$ are produced non-thermally.
The rotationally excited CH and CH$^+$ and the strong redward DIB wings are likely due to strong IR radiation from a known adjacent source; modeling of the DIB wings suggests that the carriers of those DIBs are relatively small polar molecules.

\begin{figure}[t!]
\begin{center}
\includegraphics[width=5.0in]{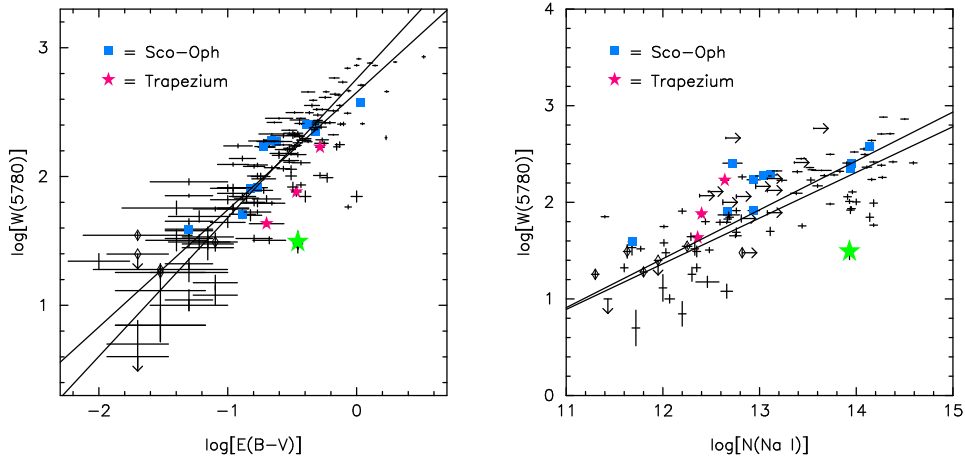}
\caption{Equivalent width of $\lambda$5780.5 DIB vs. $E(B-V)$ ({\it left}) and $N$(Na~{\sc i}) ({\it right}).
Like most of the typically strongest DIBs, the $\lambda$5780.5 DIB is very weak toward HD~62542 (large star), compared to other Galactic sight lines with similar $E(B-V)$ or $N$(Na~{\sc i}).}
\label{fig:h62542}
\end{center}
\end{figure}

{\bf HD 62542:} 
The line of sight to HD 62542 is remarkable for its very steep far-UV extinction, high column densities of CH, C$_2$, CN, and CO (for $A_{\rm V}$ $\sim$ 1), and apparent dearth of CH$^+$ and diffuse atomic gas.
Most of the interstellar material resides in a single narrow ($b$ $\sim$ 0.8 km~s$^{-1}$) component with a high molecular fraction [$f$(H$_2$) $>$ 0.8].
\cite[Cardelli et al. (1990)]{Card90} concluded that that main cloud is a small, moderately dense, mostly molecular knot whose more diffuse outer layers have been stripped away by stellar winds and shocks within the Gum Nebula.
\cite[Snow et al. (2002)]{Snow02} noted that the typically stronger DIBs (e.g., $\lambda$5780.5, $\lambda$5797.1, $\lambda$6283.8) are extremely weak, relative to $E(B-V)$, K~{\sc i}, and CH (Fig.~\ref{fig:h62542}).
Higher S/N spectra obtained by \cite[\'{A}d\'{a}mkovics et al. (2005)]{Adam05}, however, enabled detection of a number of the C$_2$-DIBs, with strengths similar to those seen for other sight lines with comparable $E(B-V)$, Na~{\sc i}, and K~{\sc i}.
The DIB strengths toward HD~62542 thus provide strong support for the picture in which the typically strongest DIBs trace diffuse, primarily atomic gas, while the generally weaker C$_2$-DIBs trace denser gas with a significant molecular fraction.

\section{DIBs and Physical Conditions}

If the behavior of the DIBs can be linked to local physical conditions ($n_{\rm e}$, $n_{\rm H}$, $T_{\rm k}$, $I_{\rm UV}$ -- as derived from various atomic and molecular species), then the DIBs could be used as independent, widely available diagnostics of those physical conditions (averaged over a given sight line).
As noted above, the $W$(5797.1)/$W$(5780.5) ratio seems to have some dependence on $I_{\rm UV}$; other such ratios may be even more sensitive, given the range in correlation slopes found for different DIBs.
It will be important, however, to compare multiple, independent diagnostics (e.g., the $n_{\rm e}$ from different X~{\sc i}/X~{\sc ii} ratios, the $I_{\rm UV}$ from chemical models and H$_2$ excitation, the $n_{\rm H}$ from C~{\sc i} fine-structure and C$_2$ rotational excitation), for a variety of environments, to accurately assess the validity and robustness of any DIB-based diagnostics.
(See also the contribution by Sonnentrucker.)


\cite[Ka\'{z}mierczak et al. (2009, 2010)]{Kaz09} noted a possible correlation between the FWHM of the $\lambda$6196.0 DIB and the C$_2$ rotational excitation temperature.
Such a correlation would be somewhat unexpected, as the $\lambda$6196.0 DIB is not a C$_2$-DIB (and so likely arises in more diffuse, atomic gas); no correlation was seen for the $\lambda$4963.9 C$_2$-DIB.
The sight line sample is small, however, and the correlation appears to be largely driven by several sight lines in Sco-Oph with large FWHM(6196.0).
Examination of a larger, more diverse sample of sight lines confirms the variations in FWHM(6196.0) (see also \cite[Galazutdinov et al. 2008]{Gala08}), but reveals no convincing trends for FWHM(6196.0) vs. $T_{\rm k}$(C$_2$) or $T_{01}$(H$_2$) (both measures of the local kinetic temperature; Welty et al., in prep.).

\section{Conclusions / Prospects}

While comparisons of DIBs with other gas-phase species will, of necessity, continue to employ correlations between DIB equivalent widths and integrated sight line column densities, the underlying complex structure of the ISM (and the relevant physical and chemical processes) should be kept in mind.
Because correlations between $W$(DIB) and $N$(X) reflect both the general increase of all interstellar species with $N$(H$_{\rm tot}$) and the multiple factors affecting the abundances of the DIB carrier and species X, attention should be given not only to $r$, but to the slope of the relationship and to the scatter and residuals relative to the best fit.
Suitable ratios can be used to remove the mutual general dependence on $N$(H$_{\rm tot}$) -- to isolate specific factors affecting the relationship.

While very few DIBs have been investigated in detail so far, a few conclusions seem fairly secure:
(1) most of the stronger DIBs arise primarily in diffuse atomic gas, but the weaker C$_2$-DIBs are associated with denser, more molecular gas;
(2) many DIBs are correlated better with $E(B-V)$ and $N$(H$_{\rm tot}$) than with any individual atomic or molecular species -- though several are well correlated with H and some of the narrower DIBs are well correlated with Na~{\sc i} and K~{\sc i};
(3) the broader DIBs generally have smaller correlation slopes (e.g., vs. H$_{\rm tot}$) than the narrower DIBs -- and may be less sensitive to $I_{\rm UV}$ and more broadly distributed spatially;
(4) strong radiation fields appear to weaken many of the DIBs (vs. H$_{\rm tot}$) in several regions (Sco-Oph, Trapezium); additional metallicity-related effects further weaken the DIBs in the Magellanic Clouds; and
(5) the $W$(5797.1)/$W$(5780.5) ratio (and possibly others) can be indicative of $I_{\rm UV}$.
Observations of the DIBs in ``unusual'' environments (e.g., toward Herschel~36 and HD~62542) can provide valuable constraints on DIB behavior.

With the advent of larger surveys (many DIBs and/or many sight lines), we may be able to use more sophisticated statistical tools (e.g., principal component analysis) to uncover significant relationships among the DIBs and to distinguish the multiple factors that affect their strengths.
Automated measurement of the DIBs -- taking into account telluric and stellar absorption and differences in the DIB profiles -- will be necessary (e.g., \cite[Puspitarini et al. 2013]{Pus13}).
Large surveys should also reveal more sight lines with unusual DIB abundances and/or properties -- resulting from extreme environmental conditions -- that may provide additional insights into the nature and behavior of the DIB carriers.
Limitations in spectral coverage, resolution, and/or S/N in those surveys may render comparisons between the DIBs and the atomic and molecular species difficult, however.

Measurement of the DIBs in sight lines with existing high-resolution UV spectra would enable more direct comparisons with local densities and radiation fields (e.g., \cite[Jenkins \& Tripp 2011]{JT11}), as well as with C~{\sc i} and S~{\sc i} (which have higher $\chi_{\rm ion}$ than Na~{\sc i} and K~{\sc i}).

Detailed models for interstellar clouds that can reproduce both the abundances and relative distributions of different species (e.g., \cite[Ruiterkamp et al. 2005]{Ruit05}) should lead to better understanding of the physical/chemical conditions where the DIBs arise.

\noindent
{\bf Acknowledgements:}
It has been a pleasure to work with (and benefit from the experience and insights of) J. Dahlstrom, S. Friedman, L. Hobbs, B. McCall, T. Oka, B. Rachford, T. Snow, P. Sonnentrucker, D. York, the others in our DIBs working group, and longtime collaborator S. Federman.
This work has been supported by the National Science Foundation, under grant AST-1238926 to the University of Chicago.




\begin{thebibliography}{}

\bibitem[\'{A}d\'{a}mkovics et al. (2005)]{Adam05}
{\'{A}d\'{a}mkovics, M., Blake G.A., McCall, B.J.} 2005, \textit{ApJ}, 625, 857

\bibitem[Cardelli et al. (1990)]{Card90}
{Cardelli, J.A., Edgar, R.J., Savage, B.D., Suntzeff, N.B.} 1990, \textit{ApJ}, 362, 551

\bibitem[Cordiner et al. (2006)]{Cord06}
{Cordiner, M.A., Fossey, S.J., Smith, A.M., Sarre, P.J.} 2006, \textit{Faraday Disc}, 133, 403

\bibitem[Cordiner et al. (2013)]{Cord13}
{Cordiner, M.A., Fossey, S.J., Smith, A.M., Sarre, P.J.} 2013, \textit{ApJL}, 764, L10

\bibitem[Cox et al. (2006)]{Cox06}
{Cox, N.L.J., Cordiner, M.A., Cami, J. et al.} 2006, \textit{A\&A}, 447, 991

\bibitem[Cox et al. (2007)]{Cox07}
{Cox, N.L.J., Cordiner, M.A., Ehrenfreund, P. et al.} 2007, \textit{A\&A}, 470, 941

\bibitem[Dahlstrom et al. (2013)]{Dahl13}
{Dahlstrom, J., York, D.G., Welty, D.E. et al.} 2013, \textit{ApJ}, in press (arXiv:1305.3003)

\bibitem[Friedman et al. (2011)]{Fried11}
{Friedman, S.D., York, D.G., McCall, B.J. et al.} 2011, \textit{ApJ}, 727, 33

\bibitem[Galazutdinov et al. (2008)]{Gala08}
{Galazutdinov, G.A., Lo Curto, G., Kre{\l}owski, J.} 2008, \textit{MNRAS}, 386, 2003

\bibitem[Galazutdinov et al. (2004)]{Gala04}
{Galazutdinov, G.A., Manic\`{o}, G., Pirronello, V., Kre{\l}owski, J.} 2004, \textit{MNRAS}, 355, 169

\bibitem[Heiles (1997)]{Heil97}
{Heiles, C.} 1997, \textit{ApJ}, 481, 193

\bibitem[Herbig (1993)]{Herb93}
{Herbig, G.H.} 1993, \textit{ApJ}, 407, 142

\bibitem[Hobbs et al. (2008)]{Hob08}
{Hobbs, L.M., York, D.G., Snow, T.P. et al.} 2008, \textit{ApJ}, 680, 1256

\bibitem[Hobbs et al. (2009)]{Hob09}
{Hobbs, L.M., York, D.G., Thorburn, J.A. et al.} 2009, \textit{ApJ}, 705, 32

\bibitem[Indriolo \& McCall (2012)]{Ind12}
{Indriolo, N., \& McCall, B.J.} 2012, \textit{ApJ}, 745, 91

\bibitem[Jenkins (2009)]{Jenk09}
{Jenkins, E.B.} 2009, \textit{ApJ}, 700, 1299

\bibitem[Jenkins \& Tripp (2011)]{JT11}
{Jenkins, E.B., \& Tripp, T.M.} 2011, \textit{ApJ}, 734, 65

\bibitem[Ka\'{z}mierczak et al. (2009)]{Kaz09}
{Ka\'{z}mierczak, M., Gnaci\'{n}ski, P., Schmidt, M.R. et al.} 2009, \textit{A\&A}, 498, 785

\bibitem[Ka\'{z}mierczak et al. (2010)]{Kaz10}
{Ka\'{z}mierczak, M., Schmidt, M.R., Galazutdinov, G.A. et al.} 2010, \textit{MNRAS}, 408, 1590

\bibitem[Kre{\l}owski et al. (1998)]{Krel98}
{Kre{\l}owski, J., Galazutdinov, G.A., Musaev, F.A.} 1998, \textit{ApJ}, 493, 217

\bibitem[Kre{\l}owski et al. (1999)]{Krel99}
{Kre{\l}owski, J., Ehrenfreund, P., Foing, B.H. et al.} 1999, \textit{A\&A}, 347, 235

\bibitem[Lauroesch (2007)]{Laur07}
{Lauroesch, J.T.} 2007, in: M. Haverkorn \& W.M. Goss (eds.), ASP Conf. Ser. 365, \textit{SINS -- Small Ionized and Neutral Structures in the Diffuse Interstellar Medium} (San Francisco: Astron. Soc. Pacific), p. 40

\bibitem[Liszt (2003)]{Liszt03}
{Liszt, H.S.} 2003, \textit{A\&A}, 398, 621

\bibitem[Oka et al. (2013)]{Oka13}
{Oka, T., Welty, D.E., Johnson, S. et al.} 2013, \textit{ApJ}, in press (arXiv:1304.2842)

\bibitem[Pan et al. (2005)]{Pan05}
{Pan, K., Federman, S.R., Sheffer, Y., Andersson, B.-G.} 2005, \textit{ApJ}, 633, 986

\bibitem[Price et al. (2001)]{Price01}
{Price, R.J., Crawford, I.A., Barlow, M.J., Howarth, I.D.} 2001, \textit{MNRAS}, 328, 555

\bibitem[Puspitarini et al. (2013)]{Pus13}
{Puspitarini, L., Lallement, R., Chen, H.-C.} 2013, \textit{A\&A}, 555, A25

\bibitem[Raimond et al. (2012)]{Rai12}
{Raimond, S., Lallement, R., Vergely, J. L. et al.} 2012, \textit{A\&A}, 544, A136

\bibitem[Ruiterkamp et al. (2005)]{Ruit05}
{Ruiterkamp, R., Cox, N.L.J., Spaans, M. et al.} 2005, \textit{A\&A}, 432, 515

\bibitem[Sheffer et al. (2008)]{Shef08}
{Sheffer, Y., Rogers, M., Federman, S.R. et al.} 2008, \textit{ApJ}, 687, 1075

\bibitem[Snow et al. (2002)]{Snow02}
{Snow, T.P., Welty, D.E., Thorburn, J. et al.} 2002, \textit{ApJ}, 573, 670


\bibitem[Thorburn et al. (2003)]{Thor03}
{Thorburn, J.A., Hobbs, L.M., McCall, B.J. et al.} 2003, \textit{ApJ}, 584, 339

\bibitem[Vos et al. (2011)]{Vos11}
{Vos, D.A.I., Cox, N.L.J., Kaper, L. et al.} 2011, \textit{A\&A}, 533, A129

\bibitem[Weingartner \& Draine (2001)]{Wein01}
{Weingartner, J.C., \& Draine, B.T.} 2001, \textit{ApJ}, 563, 842

\bibitem[Welty (2007)]{Welty07}
{Welty, D.E.} 2007, \textit{ApJ}, 668, 1012

\bibitem[Welty \& Crowther (2010)]{WC10}
{Welty, D.E., \& Crowther, P.A.} 2010, \textit{MNRAS}, 404, 1321

\bibitem[Welty et al. (2006)]{Welty06}
{Welty, D.E., Federman, S.R., Gredel, R. et al.} 2006, \textit{ApJS}, 165, 138

\bibitem[Welty \& Hobbs (2001)]{WH01}
{Welty, D.E., \& Hobbs, L.M.} 2001, \textit{ApJS}, 133, 345

\bibitem[Welty et al. (2003)]{WHM03}
{Welty, D.E., Hobbs, L.M., Morton, D.C.} 2003, \textit{ApJS}, 147, 61

\bibitem[Welty et al. (1996)]{WMH96}
{Welty, D.E., Morton, D.C., Hobbs, L.M.} 1996, \textit{ApJS}, 106, 533

\bibitem[Weselak et al. (2010)]{Wes10}
{Weselak, T., Galazutdinov, G.A., Han, I., Kre{\l}owski, J.} 2010, \textit{MNRAS}, 401, 1308

\bibitem[Weselak et al. (2004)]{Wes04}
{Weselak, T., Galazutdinov, G.A., Musaev, F.A., Kre{\l}owski, J.} 2004, \textit{A\&A}, 414, 949

\bibitem[Weselak et al. (2008)]{Wes08}
{Weselak, T., Galazutdinov, G.A., Musaev, F.A., Kre{\l}owski, J.} 2008, \textit{A\&A}, 484, 381

\end{thebibliography}
\end{document}